\documentclass[aps,superscriptaddress,amsmath,amssymb,twocolumn,showpacs,floatfix,english,noshowpacs]{revtex4}

\usepackage{url}
\usepackage{bm,bbm}
\usepackage{graphicx}
\usepackage{color}
\usepackage[colorlinks=true, urlcolor=blue, linkcolor=blue, citecolor=blue, pdftex]{hyperref}
\usepackage[english]{babel}

\usepackage{soul}

\newcommand{\dagga}{{\phantom{\dagger}}}

\begin{document}

\title{Spectral signatures of fractionalization in the frustrated Heisenberg model on the square lattice}

\author{Francesco Ferrari}
\affiliation{SISSA-International School for Advanced Studies, Via Bonomea 265, I-34136 Trieste, Italy}
\author{Federico Becca}
\affiliation{Democritos National Simulation Center, Istituto Officina dei Materiali del CNR and
SISSA-International School for Advanced Studies, Via Bonomea 265, I-34136 Trieste, Italy}

\date{\today}

\begin{abstract}
We employ a variational Monte Carlo approach to efficiently obtain the dynamical structure factor for the spin-1/2 $J_1-J_2$ Heisenberg model on the 
square lattice. Upon increasing the frustrating ratio $J_2/J_1$, the ground state undergoes a continuous transition from a N\'eel antiferromagnet to a 
$\mathbb{Z}_{2}$ gapless spin liquid. We identify the characteristic spectral features in both phases and highlight the existence of a broad continuum 
of excitations in the proximity of the spin-liquid phase. The magnon branch, which dominates the spectrum of the unfrustrated Heisenberg model, gradually 
loses its spectral weight, thus releasing nearly-deconfined spinons, whose signatures are visible even in the magnetically ordered state. Our results 
provide an important example on how magnons fractionalize into deconfined spinons across a quantum critical point.
\end{abstract}

\maketitle

{\it Introduction.}
The existence of fractional excitations is a phenomenon due to the strong interaction between the particles. It refers to the emergence of quasiparticle
excitations having quantum numbers that are non-integer multiples of those of the constituent particles (such as electrons)~\cite{laughlin1999}. In two
or three spatial dimensions fractionalization is generally associated with emergent gauge fields and can be described in terms of deconfinement of 
quasiparticles that are free to move, rather than forming bound states. There are several examples of fractionalization, ranging from high-energy physics 
to condensed matter, where gauge theories show transitions between confining and deconfining phases~\cite{kogut1979}. One of the most prominent examples 
is the fractional quantum Hall effect, where the quasiparticles carry fractions of the electron charge~\cite{laughlin1983}. Another important case is 
given by the spin-charge separation in one-dimensional electronic conductors: here, the electron splits into a {\it spinon}, which carries spin $S=1/2$ 
and no charge, and a {\it holon}, with $S=0$ and unit charge $e$~\cite{luttinger1963,haldane1981}. Also one-dimensional insulators exhibit 
fractionalization in the spin sector, e.g., the Heisenberg model~\cite{bethe1931}, where elementary excitations are $S=1/2$ free spinons and not 
conventional $S=1$ spin waves~\cite{faddeev1981}. 

\begin{figure*}
\includegraphics[width=\columnwidth]{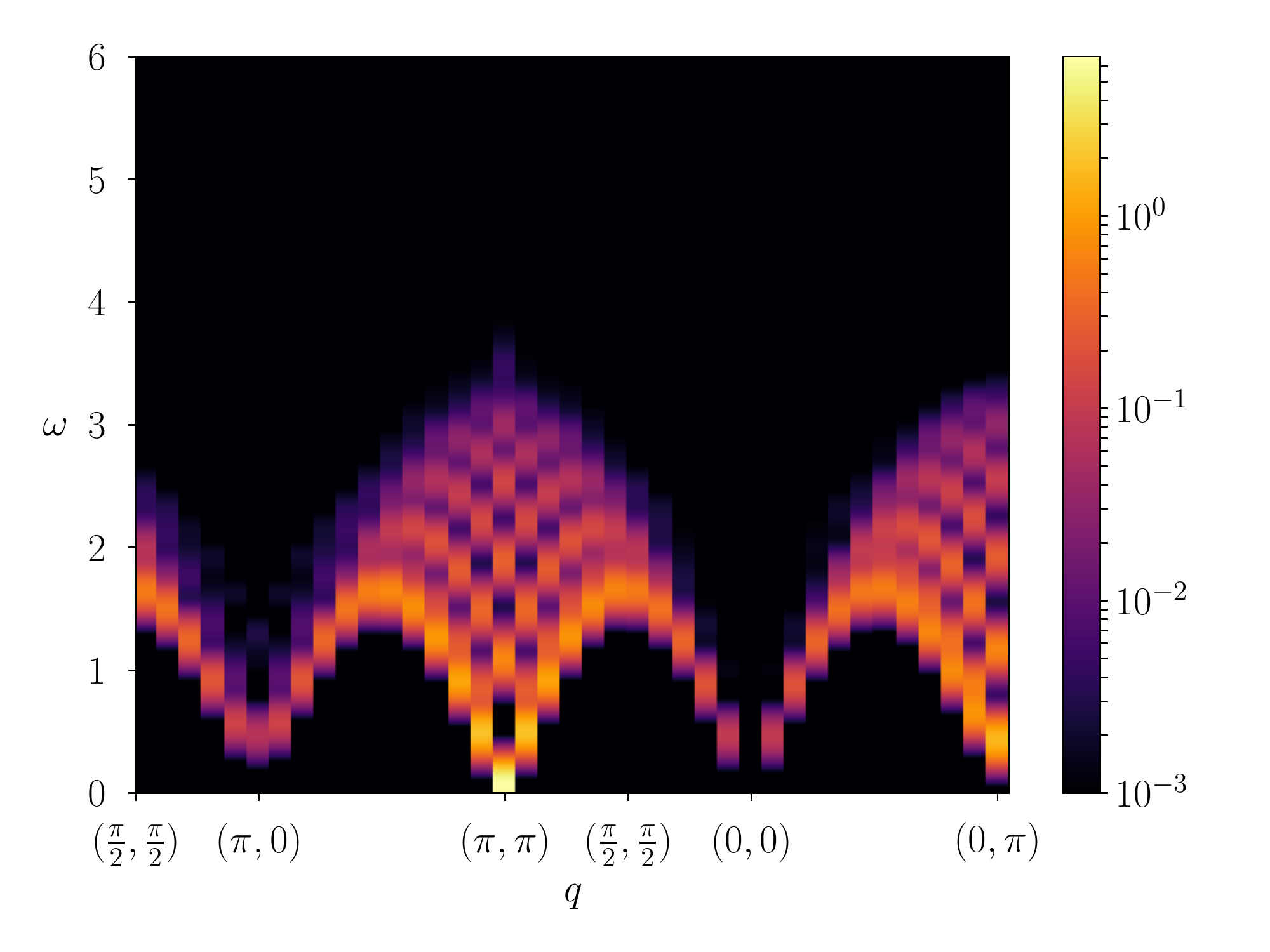}
\includegraphics[width=\columnwidth]{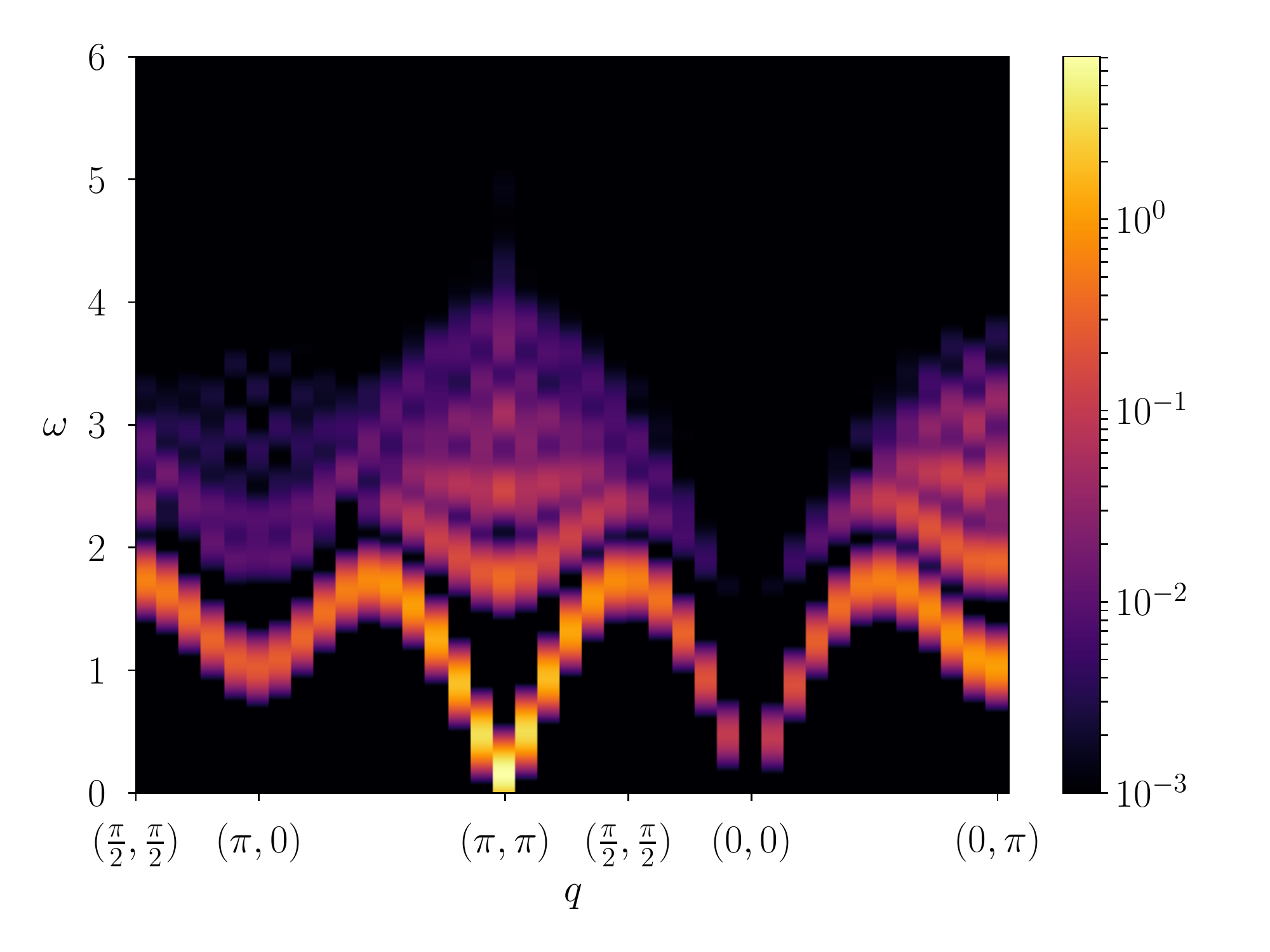}
\includegraphics[width=\columnwidth]{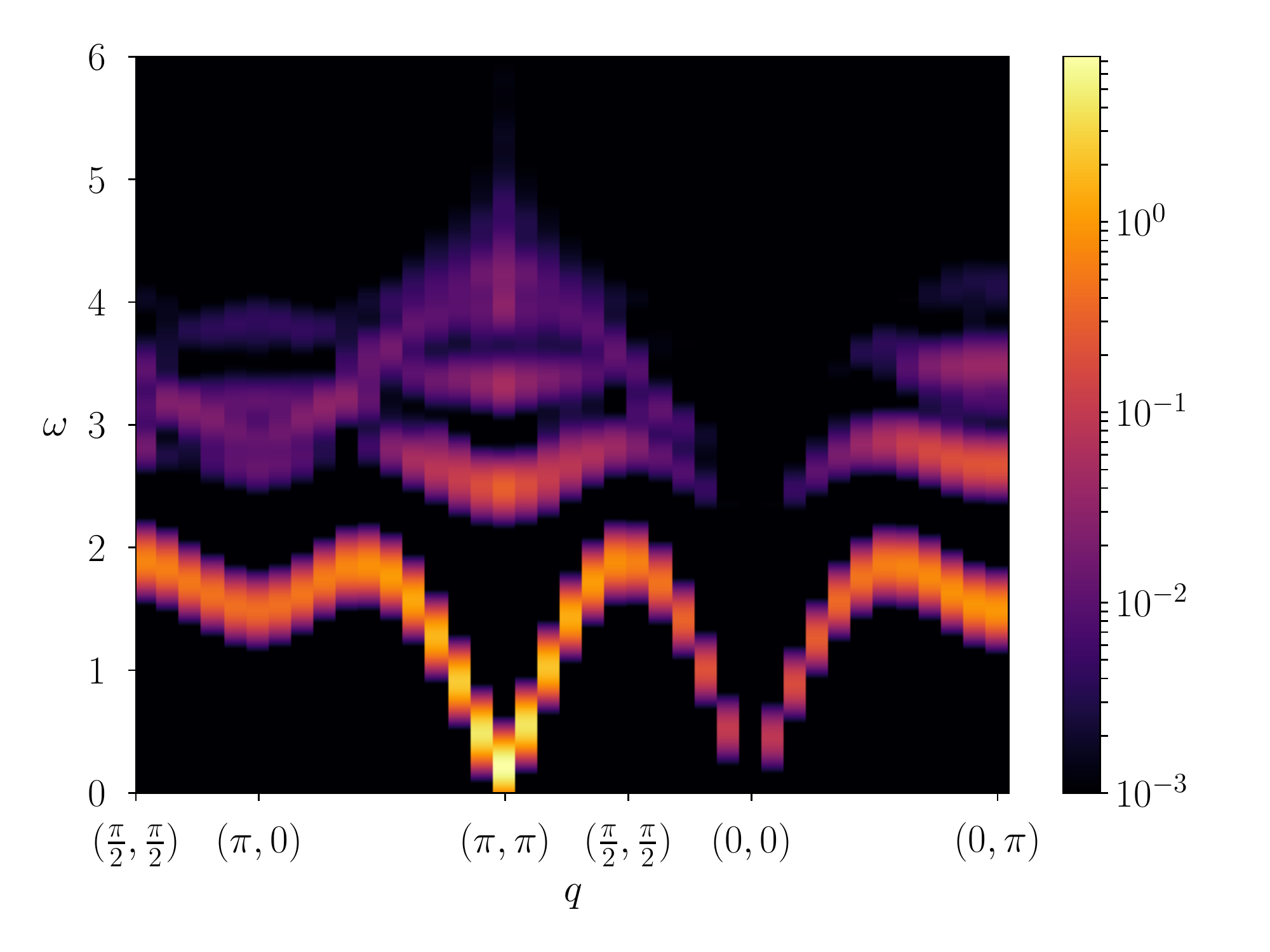}
\includegraphics[width=\columnwidth]{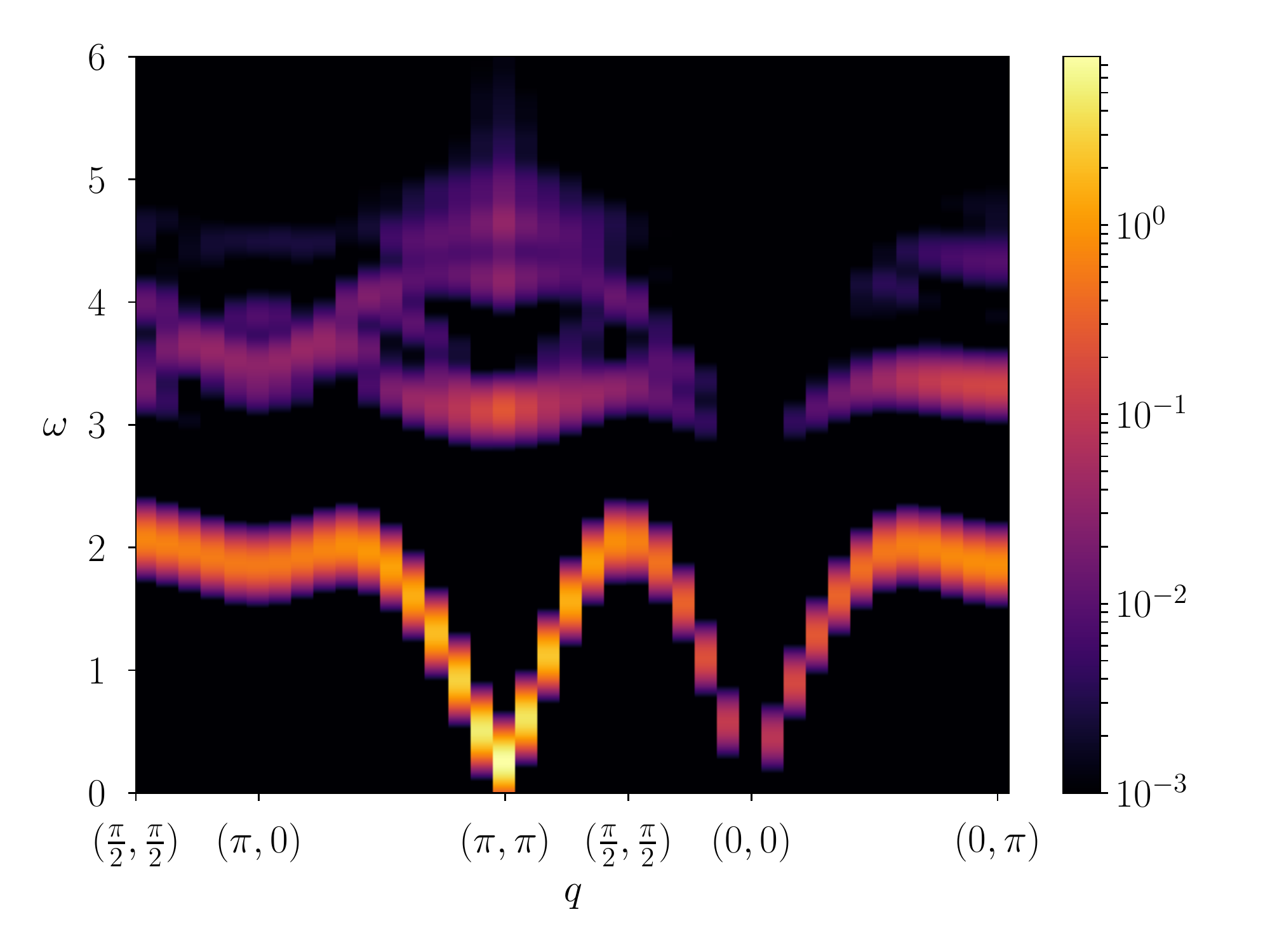}
\caption{
\label{fig:Jperp} Dynamical spin structure factor of the spatially anisotropic Heisenberg model with $J_2=0$. Different values of the inter-chain 
couplings are reported: $J_1^{\perp}/J_1^{\parallel}=0.1$ (upper-left), $0.3$ (upper-right), $0.5$ (lower-left), and $0.7$ (lower-right). The square
cluster contains $N=22 \times 22$ sites. Spectral functions have been convoluted with normalized Gaussians with standard deviation $0.1J_1^{\parallel}$.}
\end{figure*}

In recent years, motivated by the discovery of high-temperature superconductors, there has been an increasing effort to understand whether spin-charge 
separation is also possible in two-dimensional Mott insulators. In fact, in two dimensions, magnetic order is likely to develop in the ground state, 
entailing conventional spin-wave excitations (i.e., $S=1$ {\it magnons})~\cite{anderson1952}. In this regard, frustrated systems are ideal candidates,
since the competition among different super-exchange couplings may prevent the system from developing long-range order~\cite{milabook,savary2017}. There 
have been several attempts to obtain fractional excitations in two-dimensional antiferromagnets, ranging from the earliest approaches based upon the 
resonating-valence bond (RVB) theory~\cite{baskaran1988,affleck1988,read1991} to alternative theoretical frameworks based on boson-vortex dualities 
and $\mathbb{Z}_{2}$ gauge theory~\cite{balents1999,senthil2000}. Examples of deconfined excitations have been detected in the Kitaev 
model~\cite{kitaev2006} or its generalization including Heisenberg terms~\cite{knolle2014,gohlke2017}. Fractionalization has been also invoked to 
describe continuous phase transitions, where, in contrast to the conventional Landau-Ginzburg-Wilson paradigm, characterized by critical fluctuation 
of the order parameter, the critical theory contains emergent gauge fields and deconfined quasiparticles with fractional quantum 
numbers~\cite{senthil2004a,senthil2004b}. In this context, dynamical signatures of the fractionalized excitations have been recently considered within 
a spin model with ring-exchange interactions~\cite{ma2018}. Furthermore, the possibility for a coexistence of nearly-free spinons and conventional 
$S=1$ magnon excitations has been suggested by recent neutron-scattering experiments on Cu(DCOO)$_2 \cdot$4D$_2$O, which have revealed the presence of 
a very broad spectrum around the wave vector $q=(\pi,0)$ [and $(0,\pi)$] together with a strong magnon peak around $q=(\pi,\pi)$~\cite{dallapiazza2015}. 
The experimental results were combined with a theoretical analysis based upon variational wave functions and the {\it unfrustrated} Heisenberg model. 
However, their theoretical description was not fully satisfactory, since the magnon branch in the whole Brillouin zone was recovered by a variational 
state that includes magnetic order, while a broad continuum of deconfined excitations was obtained when using a spin-liquid (i.e., RVB) wave function. 
Still, an intriguing interpretation of these results involves the existence of nearly-deconfined spinons~\cite{shao2017}, while a more conventional one 
resorts to magnons with a strong attraction at short length scales~\cite{powalski2018}.

In this paper, we report variational calculations for the frustrated $J_1-J_2$ spin-1/2 Heisenberg model on the square lattice. Different numerical 
approaches suggested the existence of a quantum critical point at $J_2/J_1 \approx 0.5$, separating an ordered N\'eel antiferromagnet and a non-magnetic 
phase~\cite{jiang2012,hu2013,gong2014,wang2016,poilblanc2017,wang2017,Haghshenas2018}, whose precise nature is still under debate. By using a recently
developed variational technique that can deal with low-energy states with given momentum $q$~\cite{li2010,ferrari2018}, we report the evolution of the 
dynamical structure factor from the N\'eel to a gapless $\mathbb{Z}_{2}$ spin-liquid phase, which is obtained within our approach. The gradual softening 
and broadening of the spectral signal around $q=(\pi,0)$ and $(0,\pi)$ when approaching the critical point represents an important hallmark of the spin 
model. A gapless continuum of free spinons in the spin-liquid phase is ascribed to four Dirac points at $q=(\pm \pi/2,\pm \pi/2)$. Moreover, our results
suggest the possibility that nearly-deconfined spinons already exist within the N\'eel phase close to the critical point. This work provides the evidence 
that the dynamical signatures of fractionalization observed for the sign-free model of Ref.~\cite{ma2018} is a generic feature of continuous quantum 
phase transitions in frustrated spin models.

{\it Model and method.}
We consider the $J_1-J_2$ Heisenberg model on the square lattice with $N=L \times L$ sites:
\begin{align}
{\cal H} = J_1^{\parallel} \sum_{i} {\bf S}_i \cdot {\bf S}_{i+y} + J_1^{\perp} \sum_{i} {\bf S}_i \cdot {\bf S}_{i+x} \nonumber \\
+ J_2 \sum_{i} \left ( {\bf S}_i \cdot {\bf S}_{i+x+y} + {\bf S}_i \cdot {\bf S}_{i+x-y} \right),
\end{align}
where ${\bf S}_{i}=(S^x_{i},S^y_{i},S^z_{i})$ is the spin-1/2 operator on the site $i$; the Hamiltonian contains both nearest-neighbor terms along
$x$ and $y$ spatial directions ($J_1^{\perp}$ and $J_1^{\parallel}$, respectively) and next-nearest-neighbor ones along $x+y$ and $x-y$ ($J_2$).
Here, we consider $J_1^{\perp} \ne J_1^{\parallel}$ only when $J_2=0$, while we take $J_1^{\perp}=J_1^{\parallel}=J_1$ when $J_2>0$. The central aim 
of this work is to assess the dynamical structure factor at zero temperature, defined by: 
\begin{equation}\label{eq:dsf}
S^{a}(q,\omega) = \sum_{\alpha} |\langle \Upsilon_{\alpha}^q | S^{a}_q | \Upsilon_0 \rangle|^2 \delta(\omega-E_{\alpha}^q+E_0),
\end{equation}
where $|\Upsilon_0\rangle$ is the ground state of the system with energy $E_0$, $\{|\Upsilon_{\alpha}^q \rangle\}$ are the excited states with momentum 
$q$ (relative to the ground state) and energy $E_{\alpha}^q$, and $S^{a}_q$ is the Fourier transform of the spin operator $S^a_i$. 

For generic values of the frustrating ratio $J_2/J_1$, there are no exact methods that allow us to evaluate the dynamical structure factor. Therefore, 
we resort to considering a suitable approximation, by employing Gutzwiller projected fermionic wave functions to construct accurate variational states 
to describe both the ground state and low-energy excitations. In particular, we consider an auxiliary superconducting (BCS) Hamiltonian:
\begin{eqnarray}                                                                                                           
{\cal H}_{0} &=& \sum_{i,j,\sigma} t_{i,j} c_{i,\sigma}^\dagger c_{j,\sigma}^\dagga 
              + \sum_{i,j} \Delta_{i,j} c_{i,\uparrow}^\dagger c_{j,\downarrow}^\dagger + h.c.  \nonumber \\      
             &+& \Delta_{\rm AF} \sum_{i} e^{i Q R_i} \left ( c_{i,\uparrow}^\dagger c_{i,\downarrow}^\dagga
              + c_{i,\downarrow}^\dagger c_{i,\uparrow}^\dagga \right ),
\label{eq:H0}
\end{eqnarray}
where, $c_{i,\sigma}^\dagger$ ($c_{i,\sigma}^\dagga$) creates (destroys) an electron with spin $\sigma=\pm 1/2$ on the site $i$; $t_{i,j}$ is a complex 
nearest-neighbor hopping generating a staggered magnetic flux on elementary (square) plaquettes~\cite{affleck1988} (when $J_1^{\perp} \ne J_1^{\parallel}$,
different amplitudes along $x$ and $y$ are considered); $\Delta_{i,j}=\Delta_{j,i}$ is a real singlet pairing with $d_{xy}$ symmetry~\cite{notedxy}; 
finally, $\Delta_{\rm AF}$ is an antiferromagnetic parameter pointing along $x$, with periodicity given by the pitch vector $Q=(\pi,\pi)$. A suitable 
variational wave function for the spin system is obtained by taking the ground state $|\Phi_0\rangle$ of the auxiliary Hamiltonian ${\cal H}_{0}$ and 
projecting out all the configurations with at least one empty or doubly occupied site:
\begin{equation}
|\Psi_0\rangle = \mathcal{P}_{S_z} \mathcal{J}_s \mathcal{P}_G |\Phi_0\rangle,
\end{equation}
where $\mathcal{P}_G=\prod_i(n_{i,\uparrow}-n_{i,\downarrow})^2$ ($n_{i,\sigma}= c_{i,\sigma}^\dagger c_{i,\sigma}^\dagga$ being the local electron 
density per spin $\sigma$ on site $i$) is the Gutzwiller projector. In addition, the spin-spin Jastrow factor 
$\mathcal{J}_s=\exp \left ( 1/2 \sum_{i,j} v_{i,j} S^z_i S^z_j \right )$ is also considered to include relevant spin-wave fluctuations over the staggered 
magnetization induced by $\Delta_{\rm AF}$~\cite{manousakis1991}. Finally, $\mathcal{P}_{S_z}$ is the projection onto the subspace 
with $S^z_{\rm tot}=\sum_i S^z_i=0$. Due to the correlated nature of the variational wave function, the evaluation of all physical quantities requires 
a Monte Carlo sampling. All the parameters inside ${\cal H}_{0}$, as well as the Jastrow pseudo-potential $v_{i,j}$ (which is taken to be translationally 
invariant), are optimized by mean of the stochastic reconfiguration technique, to minimize the variational energy of $|\Psi_0\rangle$~\cite{sorella2005}. 

\begin{figure}
\includegraphics[width=\columnwidth]{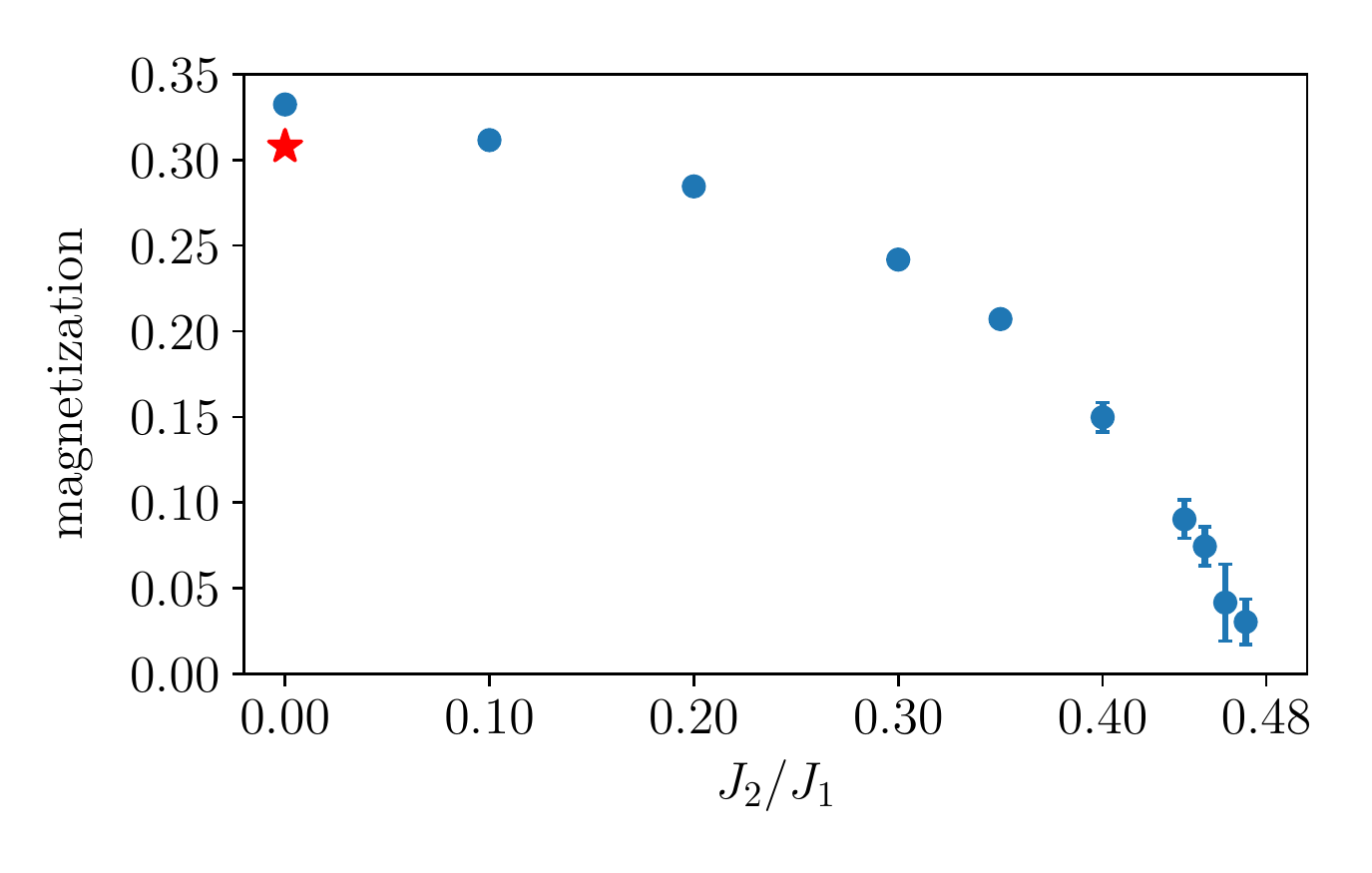}
\caption{\label{fig:magn} 
Variational results for the ground-state magnetization of the $J_1-J_2$ Heisenberg model. The results are obtained in the thermodynamic limit, by 
extrapolating the isotropic spin-spin correlations at the maximum distance in the $L \times L$ clusters with $L$ ranging from $10$ to $22$. The exact 
result for the unfrustrated Heisenberg model, obtained by quantum Monte Carlo~\cite{calandra1998,sandvik1999}, is also reported for comparison (red star).}
\end{figure}

\begin{figure*}
\includegraphics[width=\columnwidth]{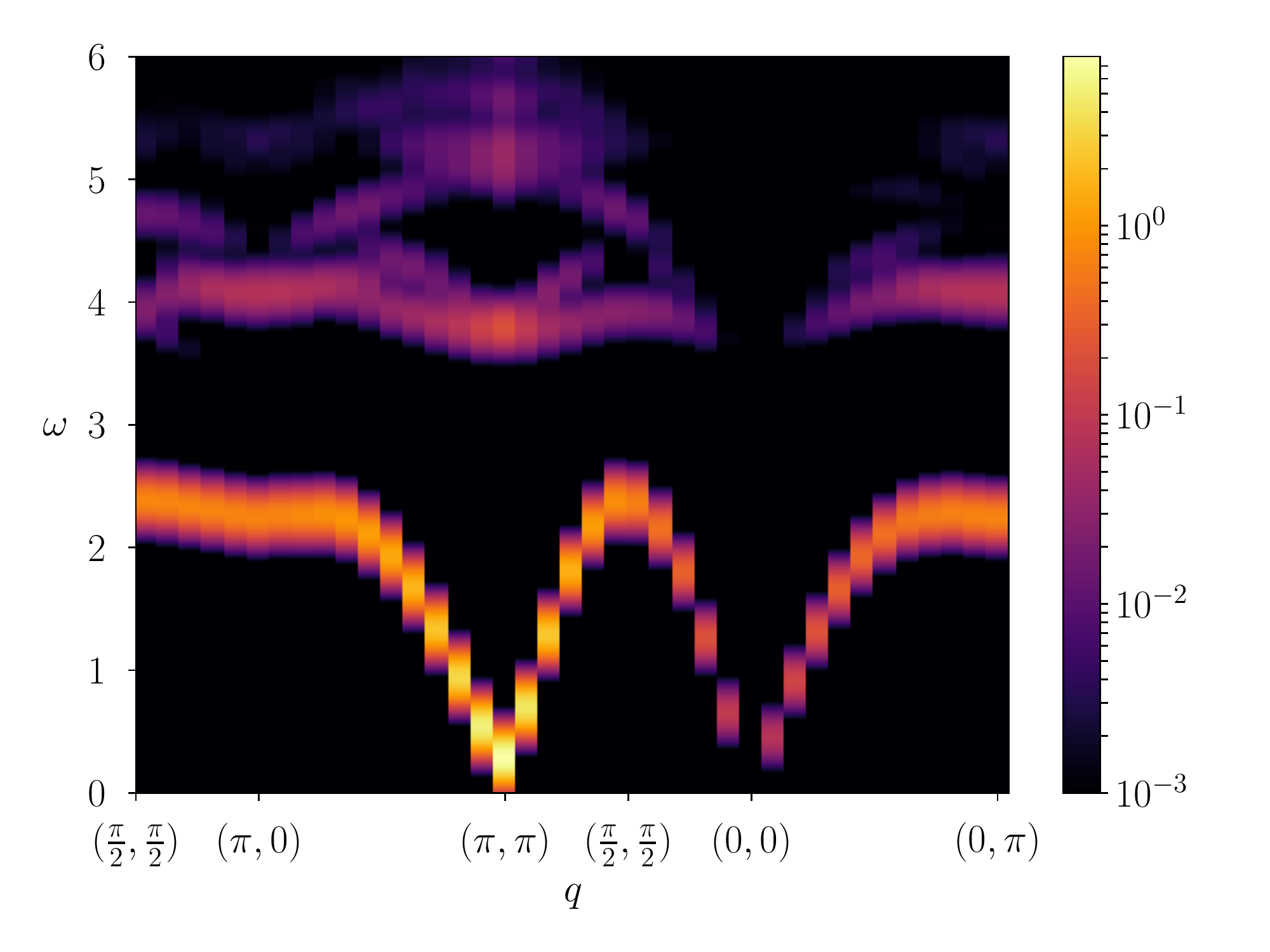}
\includegraphics[width=\columnwidth]{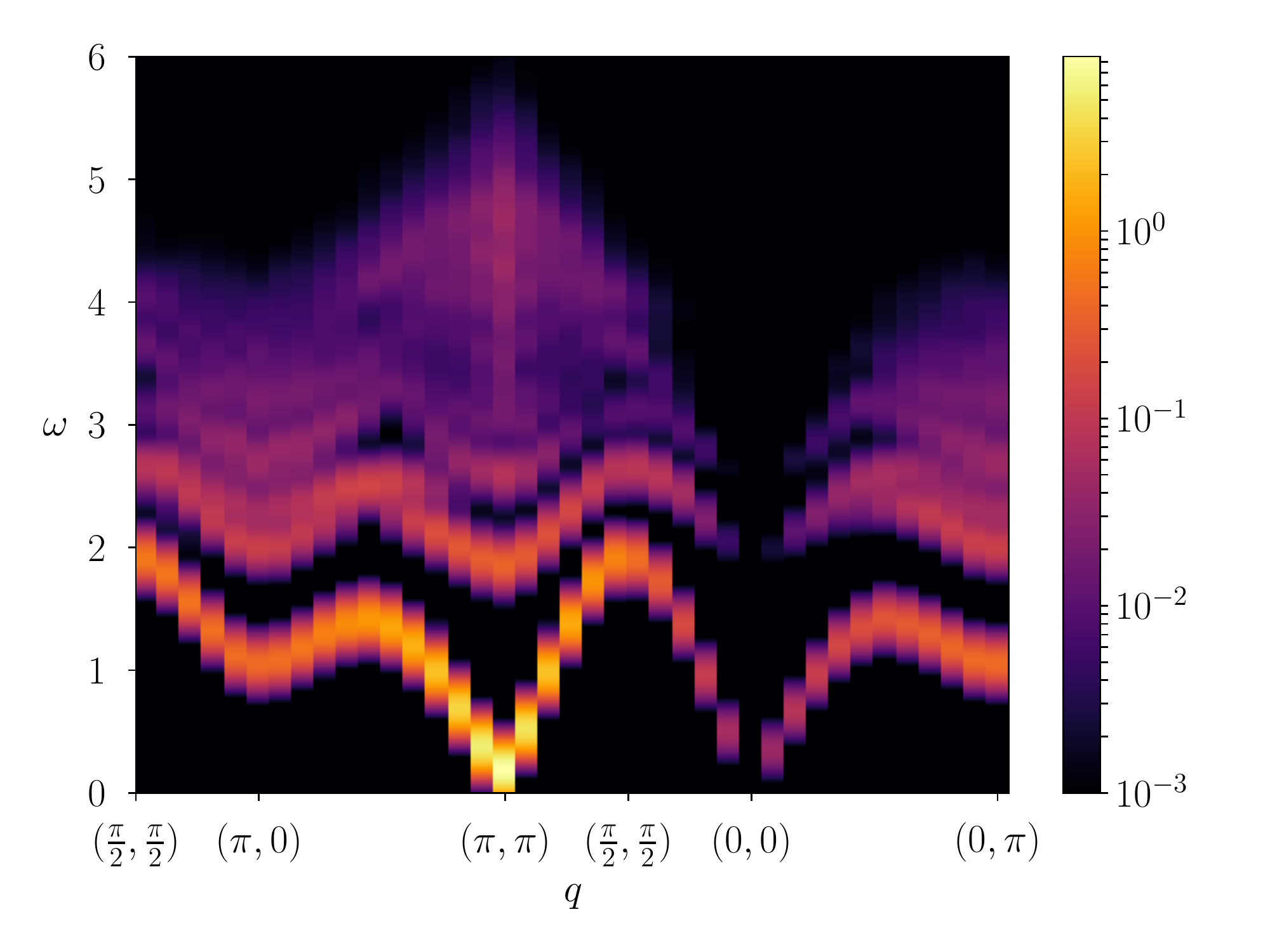}
\includegraphics[width=\columnwidth]{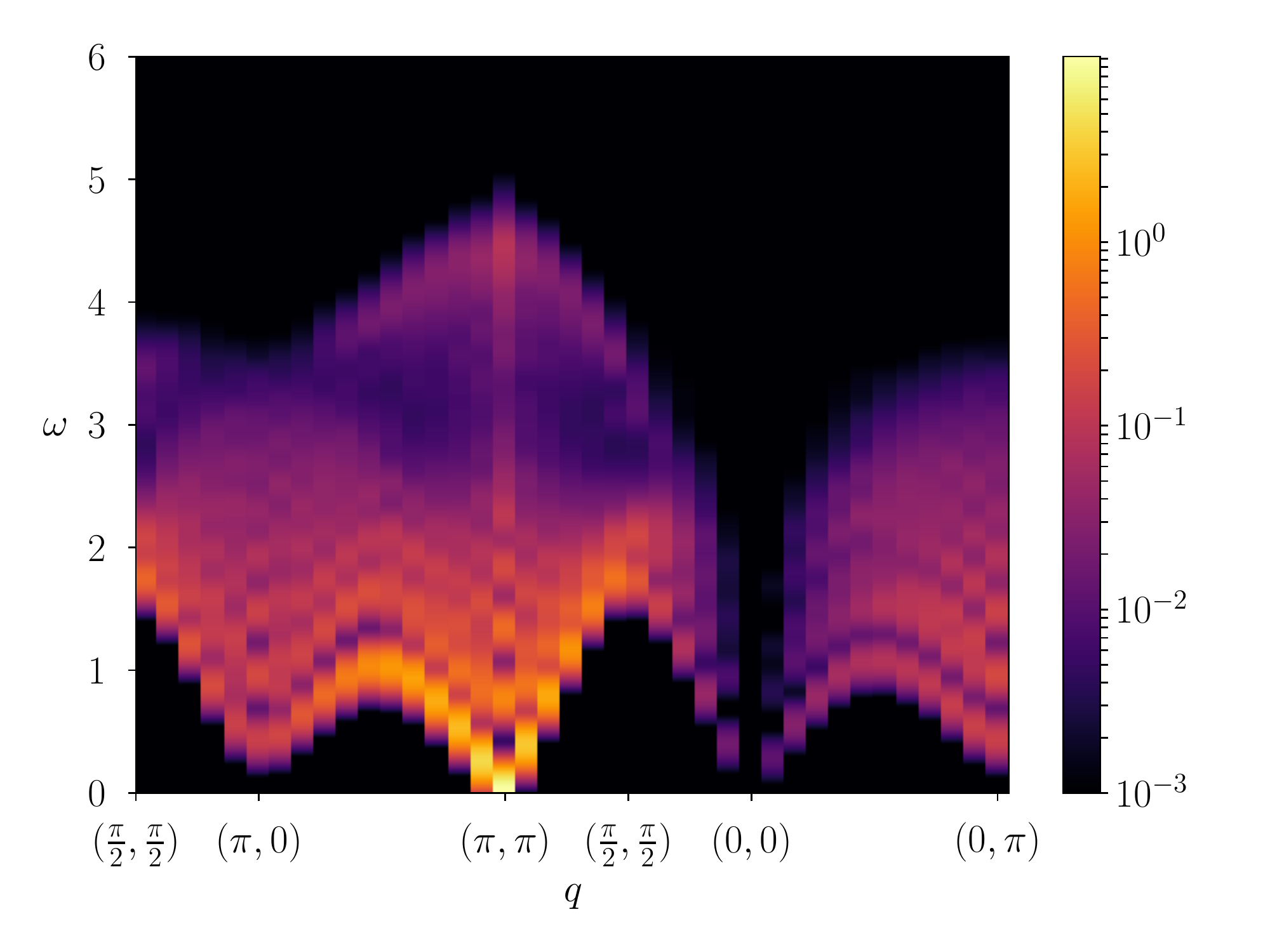}
\includegraphics[width=\columnwidth]{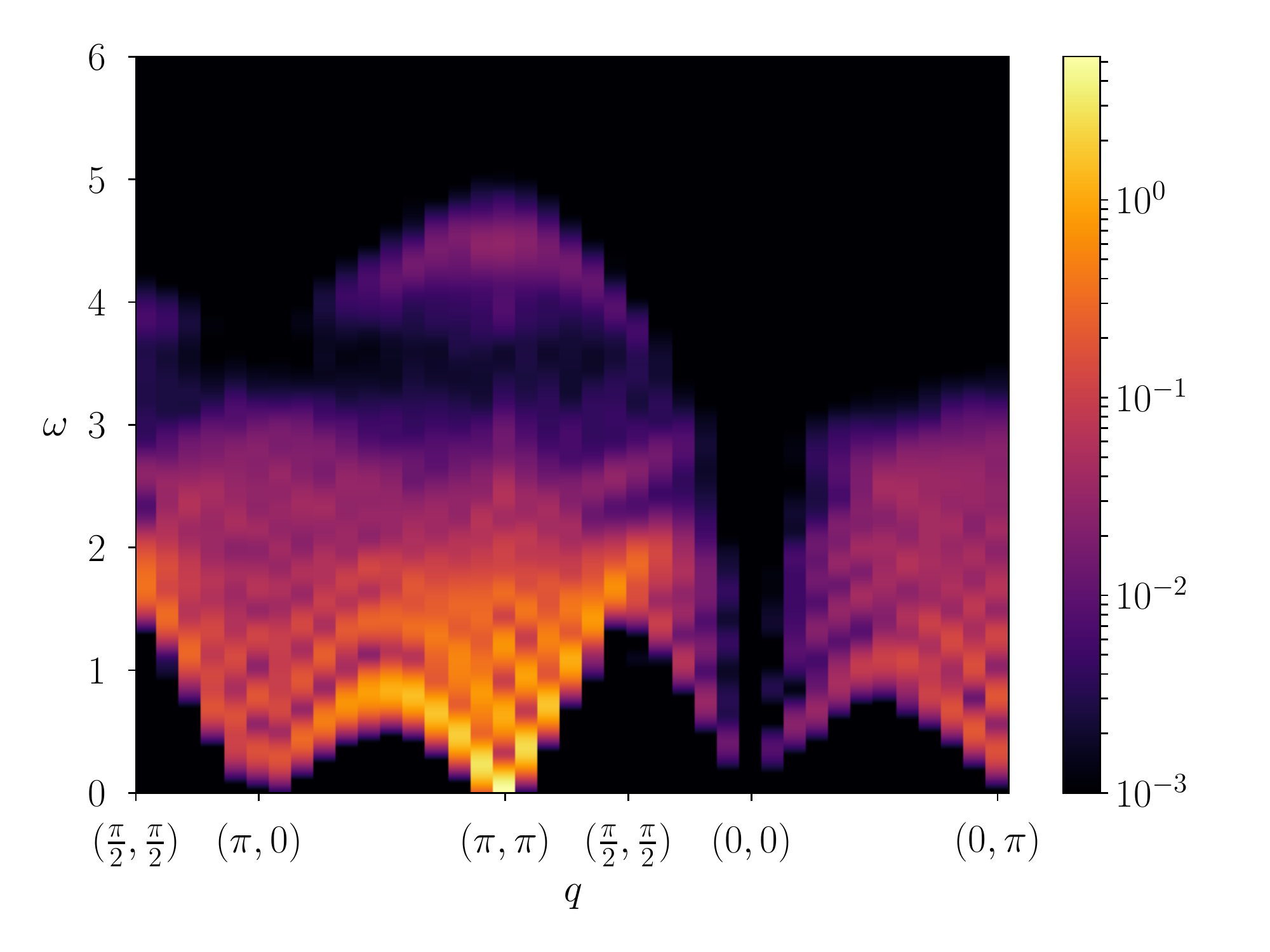}
\caption{\label{fig:J1J2} 
Dynamical spin structure factor of the $J_1-J_2$ Heisenberg model. Different values of the frustrating ratio are reported: $J_2/J_1=0$ (upper-left), 
$0.3$ (upper-right), $0.45$ (lower-left), and $0.55$ (lower-right). The antiferromagnetic parameter $\Delta_{\rm AF}$ is finite in the first three 
cases, while it is vanishing for the last one. The square cluster contains $N=22 \times 22$ sites. Spectral functions have been convoluted with 
normalized Gaussians with standard deviation $0.1J_1$.}
\end{figure*}

Following Refs.~\cite{dallapiazza2015,li2010,ferrari2018}, for  each momentum $q$, we define a (non-orthogonal) set of states, labeled by the site 
position $R$, which can be used to approximate the exact low-energy excitations:
\begin{equation}\label{eq:qRstate}
|q,R\rangle =  \mathcal{P}_{S_z} \mathcal{J}_s \mathcal{P}_G \frac{1}{\sqrt{N}} \sum_{i} e^{i q R_i} \sum_{\sigma} \sigma
c^\dagger_{i+R,\sigma}c^\dagga_{i,\sigma} |\Phi_0\rangle.
\end{equation}
By diagonalizing the Heisenberg Hamiltonian within the subspace $\{ |q,R\rangle \}$, a set of approximate excited states 
$|\Psi_n^q\rangle = \sum_R A^{n,q}_R |q,R\rangle$, with energies $\{E_{n}^q \}$, can be constructed, as described in Ref.~\cite{ferrari2018}. Then, 
the dynamical structure factor of Eq.~(\ref{eq:dsf}) is given by:
\begin{equation}\label{eq:Szz_practical}
S^{a}(q,\omega) = \sum_n |\langle \Psi_{n}^q | S^{a}_q | \Psi_0 \rangle|^2 \delta(\omega-E_{n}^q+E_0^{\rm var}),
\end{equation}
where $E_0^{\rm var}$ is the variational energy of $|\Psi_0 \rangle$. We emphasize that, within this variational approximation, the sum over excited 
states runs over at most $N$ states (instead of an exponentially large number of the exact formulation). Our approach is particularly suited to capture 
two-spinon excitations, both when they form a bound state (for example, a magnon) and when they remain free. In the following, we consider the 
$z$-component (i.e., $a \equiv z$) of the dynamical structure factor of Eq.~(\ref{eq:Szz_practical}). When $\Delta_{\rm AF}$ is finite in the auxiliary 
Hamiltonian of Eq.~(\ref{eq:H0}), the variational state breaks spin $SU(2)$ symmetry and $S^{z}(q,\omega)$ probes {\it transverse} fluctuations; instead, 
for the spin-liquid wave function, with $\Delta_{\rm AF}=0$, all the components of the dynamical structure factor give the same contribution.
We want to stress the important fact that all the quantities that define the dynamical structure factor of Eq.~(\ref{eq:Szz_practical}) can be computed
within a variational Monte Carlo scheme ({\it without} any sign problem and {\it without} any analytic continuation).

Before Gutzwiller projection, the excited states are particle-hole excitations in the BCS spectrum, which we identify as two-spinon terms. We expect this 
approach to be suited to describe excited states of deconfined phases, as for example in one-dimensional spin models~\cite{ferrari2018}. Nonetheless, 
bound states of two spinons can be also obtained, such as single-magnon excitations~\cite{dallapiazza2015}. The possibility to capture the multi-magnon 
features is more problematic. Indeed, for the unfrustrated Heisenberg model, our calculations (as well as the ones of Ref.~\cite{dallapiazza2015}) show 
that the multi-magnon continuum in the transverse signal is very weak and, most probably, cannot account for the actual results~\cite{sandvik2001}.

{\it Results.}
We start from the nearest-neighbor Heisenberg model with $J_1^{\perp} \ne J_1^{\parallel}$ and $J_2=0$. For $J_1^{\perp}=0$, the lattice is decoupled 
into $L$ copies of a one-dimensional chain with $L$ sites. In this case, there is no magnetic long-range order and the elementary excitations are spinons, 
which form a continuum of excitations in the dynamical structure factor~\cite{caux2006}. The variational Monte Carlo procedure described above gives an 
excellent description of the exact spectrum~\cite{ferrari2018}. As soon as $J_1^{\perp}$ is turned on, the ground state develops N\'eel magnetic order 
and a coherent magnon excitation settles down at low energies~\cite{affleck1994}. Within our variational approach, the optimal antiferromagnetic parameter 
$\Delta_{\rm AF}$ is finite as soon as $J_1^{\perp}>0$ (in this case, no pairing terms are considered). Our calculations show that the spinon excitations,
which characterize the spectrum of the one-dimensional Heisenberg model, are gradually pushed to a narrow region at higher energies, progressively losing 
their spectral weight. Concurrently, at low energies, a strong magnon branch sets in. The results for different values of the inter-chain super-exchange 
$J_1^{\perp}$ are shown in Fig.~\ref{fig:Jperp}. For $J_1^{\perp}/J_1^{\parallel}=0.1$, the dynamical structure factor still resembles the one of a pure 
Heisenberg chain~\cite{ferrari2018}. However, at variance with $J_1^{\perp}=0$, where the spectrum does not depend upon $q_x$, here there is already a 
sensible difference in the intensity of the lowest-energy excitations for different $q_x$: for example, at $q_y=\pi$, the strongest signal is found at 
$q_x=\pi$, due to the presence of the (weak) N\'eel order. As $J_1^{\perp}/J_1^{\parallel}$ is raised, the gap at $(\pi,0)$ and $(0,\pi)$ increases. 
In addition, the former one gains spectral weight, while the latter one loses it, until the limit of $J_1^{\perp}/J_1^{\parallel}=1$ is reached, where 
the rotational symmetry of the square lattice is recovered and the two momenta become equivalent. Remarkably, the broad continuum that characterizes the 
quasi-one-dimensional spectrum gradually disappears when approaching the two-dimensional limit. Here, the multi-magnon continuum is very weak, especially 
at low energies. In this sense, it would be tantalizing to discriminate between two possible channels for the magnon decay, one driven by a magnon-magnon 
interaction, leading to a multi-magnon decay, and another one in which the magnon splits into two spinons. While the latter one can be captured by the 
variational {\it Ansatz} of Eq.~(\ref{eq:qRstate}), the former one may go beyond our description.

The indication that deconfined spinons are released when approaching a quantum critical point comes from the analysis of the more interesting case with
$J_1^{\perp}=J_1^{\parallel}$ and frustrating $J_2$. First of all, to locate the quantum phase transition from the N\'eel to the magnetically disordered 
phase, we compute the staggered magnetization using the isotropic spin-spin correlation at maximum distance for different lattice size, and we extrapolate
to the thermodynamic limit. The results are reported in Fig.~\ref{fig:magn} and show that the magnetization drops to zero at $J_2/J_1 \approx 0.48$, as 
suggested by recent variational calculations on the spin gap~\cite{hu2013} (the exact result for the unfrustrated Heisenberg model, obtained by quantum 
Monte Carlo~\cite{calandra1998,sandvik1999}, is also reported for comparison). The disappearance of the order parameter is related to the fact that 
$\Delta_{\rm AF} \to 0$ in the auxiliary Hamiltonian~(\ref{eq:H0}). In the region where $\Delta_{\rm AF}=0$, a finite paring amplitude with $d_{xy}$ 
symmetry can be stabilized, but no energy gain is obtained by allowing translational symmetry breaking in hopping or pairing terms, thus implying that no 
valence-bond order is present. A comparison with exact results on the $6 \times 6$ lattice provides the degree of accuracy of our approach for both the 
unfrustrated and the highly-frustated cases~\cite{SupplMat}. Then, the results on the $22 \times 22$ cluster for different values of $J_2/J_1$ are reported 
in Fig.~\ref{fig:J1J2}. For weak frustration, the magnon branch is well defined in the entire Brillouin zone, including $q=(\pi,0)$, and the multi-magnon 
continuum is very weak. In the unfrustrated limit with $J_2=0$, we recover the well-known result that the lowest-energy excitation at $q=(\pi,0)$ [and 
$q=(0,\pi)$] is slightly lower than the one at $q=(\pm \pi/2,\pm \pi/2)$~\cite{dallapiazza2015,shao2017}. However, our variational approach is not able to 
capture the asymmetry between the weights of the magnon pole for these momenta. Still within the ordered phase, two principal effects are visible when 
increasing $J_2/J_1$. The first one is a gradual broadening of the spectrum, with the formation of a wide continuum close to the magnon branch. The second 
one is a clear reduction of the lowest-energy excitation at $q=(\pi,0)$, as already suggested in Ref.~\cite{hu2013}. This fact is related to the structure 
of the BCS spectrum of the auxiliary Hamiltonian (before Gutzwiller projection), which contains staggered fluxes and a finite $d_{xy}$ pairing. Here, there 
are four Dirac points at $q=(\pm \pi/2,\pm \pi/2)$, leading to two-spinon excitations that are gapless, not only for $q=(0,0)$, $(\pi,\pi)$, but also at 
$(\pi,0)$ and $(0,\pi)$. The above choice of the parameters (giving the best variational energy) corresponds to a gapless $\mathbb{Z}_2$ spin liquid, dubbed 
Z2Azz13 in Ref.~\cite{wen2002}. In presence of the Gutzwiller projection the spectrum is clearly gapless at $(\pi,\pi)$, while the gap at $(\pi,0)$ and 
$(0,\pi)$ may possess much larger size effects~\cite{SupplMat}. A similar situation appeared within the single-mode approximation of Ref.~\cite{hu2013}, 
where a variance extrapolation was necessary to prove the existence of gapless excitations at $(\pi,0)$ and $(0,\pi)$. Within the antiferromagnetic region, 
the presence of a broad continuum, which can be captured by our variational {\it Ansatz} with two-spinon excitations, suggests that nearly-deconfined 
spinons may be present even in the ordered phase~\cite{shao2017}, as evoked few years ago by field-theory arguments~\cite{balents1999}. Yet, in the 
$J_1-J_2$ model, fully deconfined spinon excitations are actually present beyond the critical point, for $0.48 \lesssim J_2/J_1 \lesssim 0.6$.

{\it Conclusions.}
In this work, we assessed the dynamical properties of a highly-frustrated (non-integrable) spin system, by using a variational Monte Carlo approach
based upon a restricted basis set of approximate excited states~\cite{li2010}. When increasing frustration, a gradual broadening of the spectrum takes
place in the whole Brillouin zone, suggesting that no coherent magnon excitations exist at the transition point~\cite{senthil2004a,senthil2004b}. 
A second important outcome is the development of gapless modes at $q=(\pi,0)$ and $q=(0,\pi)$, which also characterize the spin-liquid phase found for 
$0.48 \lesssim J_2/J_1 \lesssim 0.6$. At variance with the unfrustrated model studied in Ref.~\cite{ma2018}, where these gapless excitations are due 
to an emergent $O(4)$ symmetry (involving both N\'eel and valence-bond-solid order parameters), here they originate from the four Dirac points at 
$q=(\pm \pi/2,\pm \pi/2)$ in the spinon spectrum of the $\mathbb{Z}_2$ spin liquid. Finally, for weak frustrations, the concomitant existence of a 
magnon pole along with the broad continuum captured by two-spinon excitations hints the possibility to have an unconventional antiferromagnetic state, 
where magnons and nearly-deconfined spinons coexist~\cite{dallapiazza2015,shao2017}.

{\it Acknowledgements.} We thank A. Parola and S. Sorella for important suggestions and fruitful conversations. We also acknowledge A. Sandvik, H. Shao, 
and R. Verresen for sharing their results on the Heisenberg model with us and A. Chernyshev for useful discussions. 

{\it Note added.}
During the completion of this paper, we became aware of a similar work, based upon semi-analytical techniques, which also studied spectral properties of
the frustrated Heisenberg model, supporting the possibility for a deconfined criticality when increasing $J_2/J_1$~\cite{yu2018}.

\bibliographystyle{apsrev4-1}

\begin{thebibliography}{99}
\bibitem{notedxy} The most natural choice is to take a $d_{xy}$ pairing on the second-neighbor bonds of the lattice. However, in order to have gapless
   (Dirac) points located at the commensurate momenta $q=(\pm \frac{\pi}{2}, \pm \frac{\pi}{2})$ and achieve a better accuracy, we consider a pairing 
   with $d_{xy}$ symmetry at the fifth-neighbor distance.
\end{thebibliography}

\newpage
\widetext
\section*{SUPPLEMENTAL MATERIALS}

\begin{figure*}[!htb]
\includegraphics[width=0.495\columnwidth]{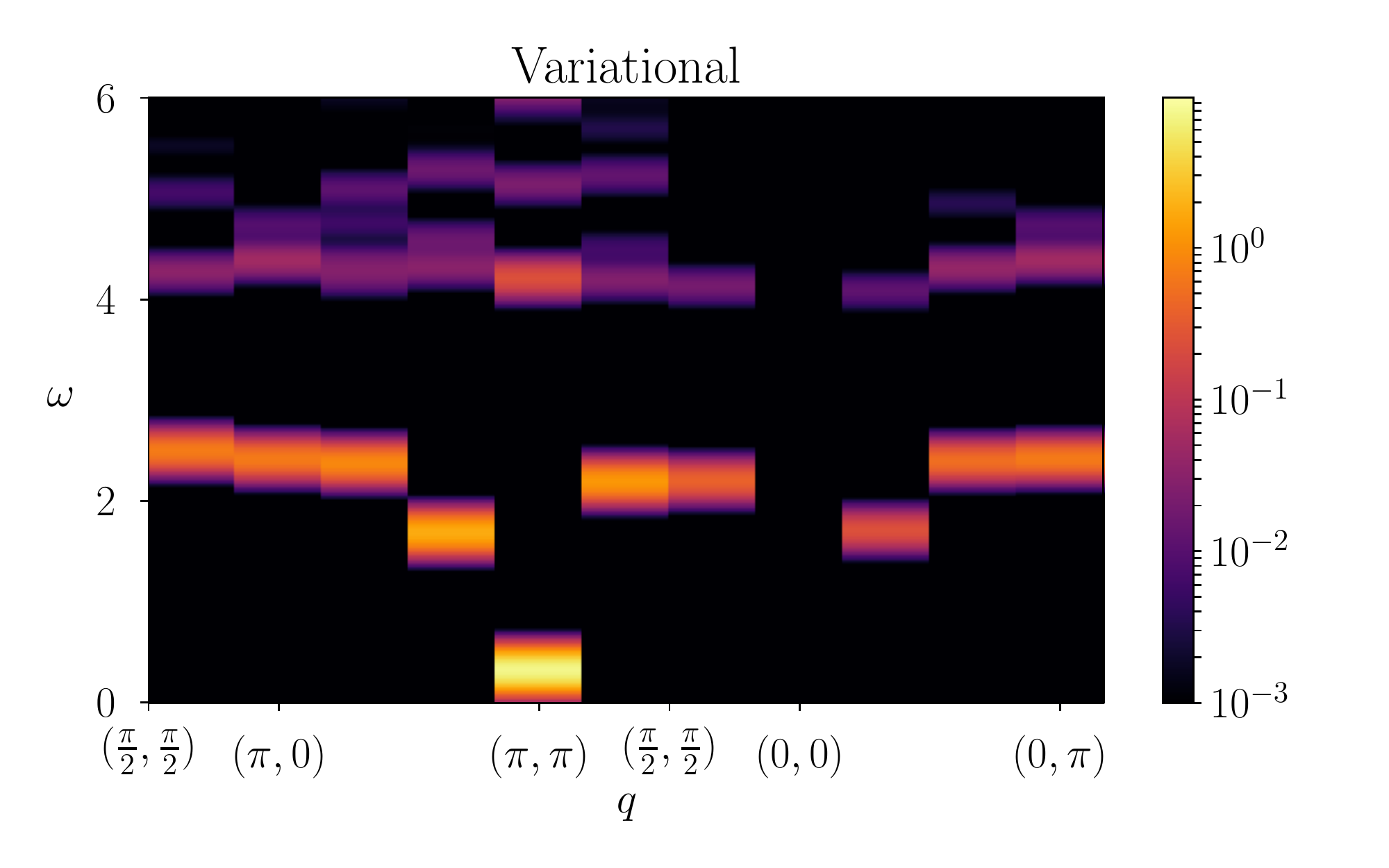}
\includegraphics[width=0.495\columnwidth]{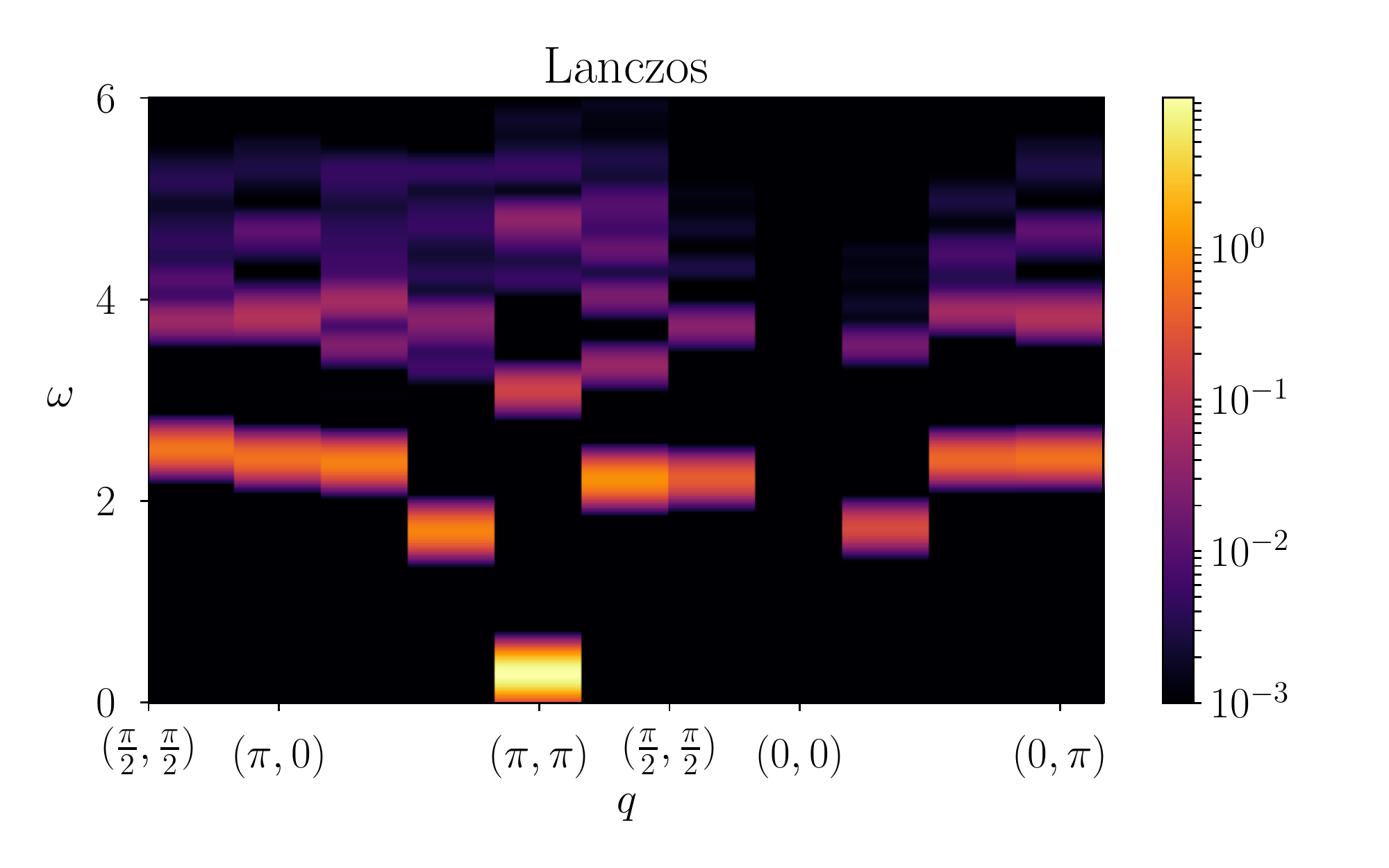}
\includegraphics[width=0.9\columnwidth]{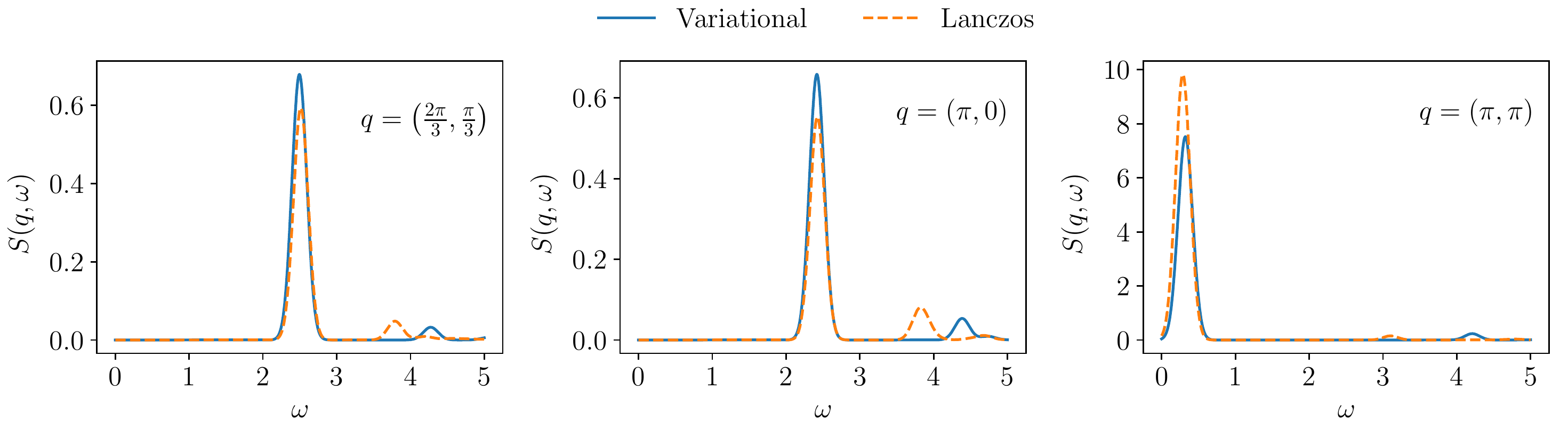}
\caption{
Dynamical spin structure factor of the unfrustrated Heisenberg model on the $6 \times 6$ square lattice. The results obtained with the variational approach 
are compared to the exact ones obtained by Lanczos diagonalization. The variational spectrum reported here corresponds to the transverse component of the 
dynamical structure factor, since the variational wave function explicitly breaks the $SU(2)$ symmetry of the Heisenberg model (due to the presence of a finite 
antiferromagnetic parameter $\Delta_{\rm AF}$). On the other hand, the spectrum obtained through Lanczos diagonalization is symmetric under $SU(2)$ rotations,
since no spontaneous symmetry breaking is possible on finite clusters. \textbf{Upper panels:} Color maps of the dynamical structure factor along a given path 
in the Brillouin zone. \textbf{Lower panels:} Dynamical structure factor for three selected momenta.}
\end{figure*}

\begin{figure*}[!htb]
\includegraphics[width=0.495\columnwidth]{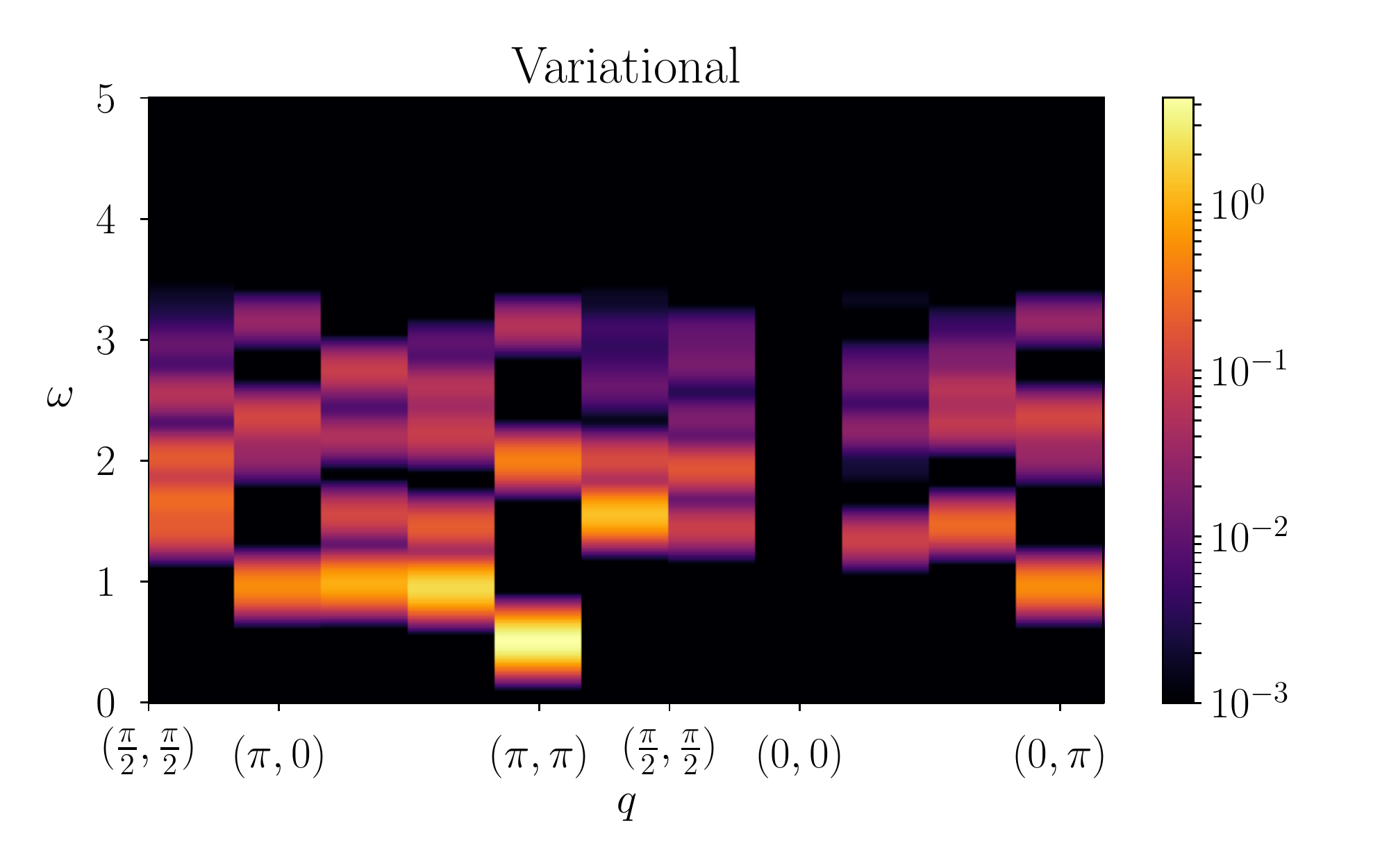}
\includegraphics[width=0.495\columnwidth]{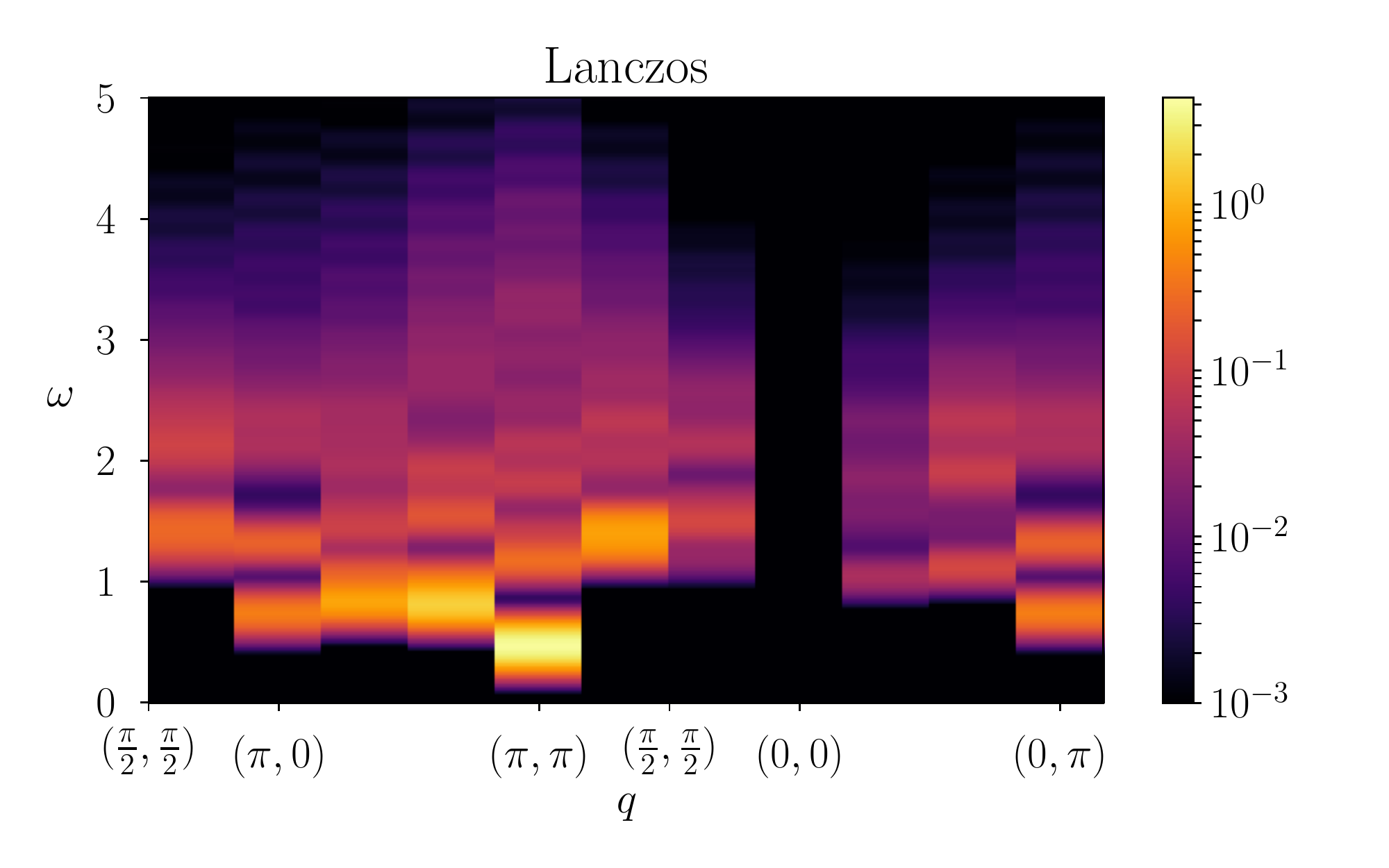}
\includegraphics[width=0.9\columnwidth]{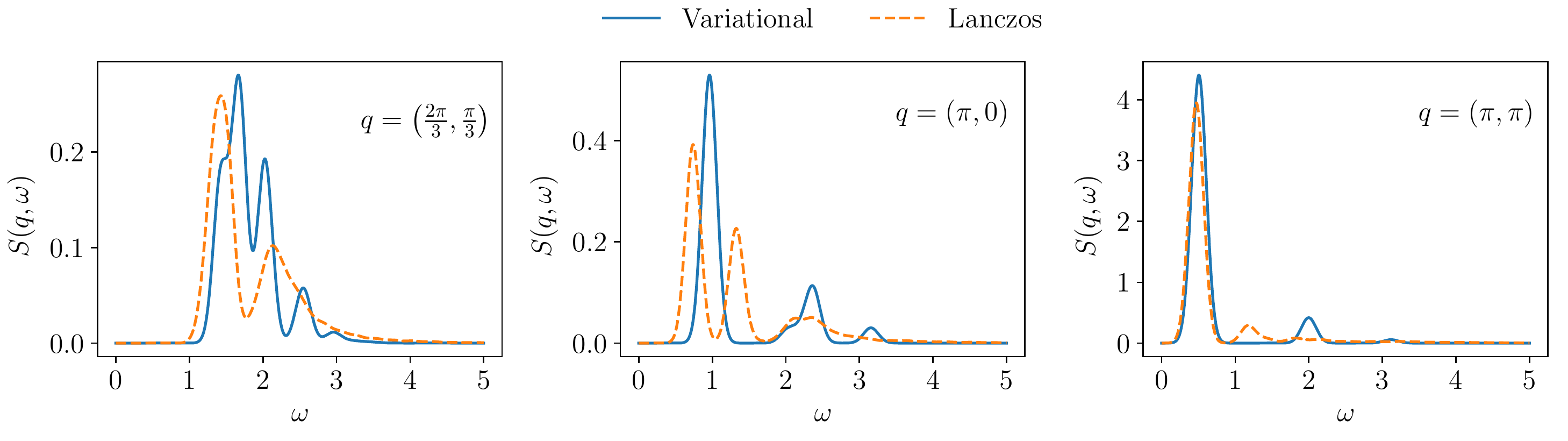}
\caption{
Dynamical spin structure factor of the $J_1-J_2$ Heisenberg model with $J_2/J_1=0.5$ on the $6 \times 6$ square lattice. The results obtained with the 
variational approach are compared to the exact ones obtained by Lanczos diagonalization. \textbf{Upper panels:} Color maps of the dynamical structure factor 
along a given path in the Brillouin zone. \textbf{Lower panels:} Dynamical structure factor for three selected momenta.}
\end{figure*}

\begin{figure*}[!htb]
\includegraphics[width=0.5\columnwidth]{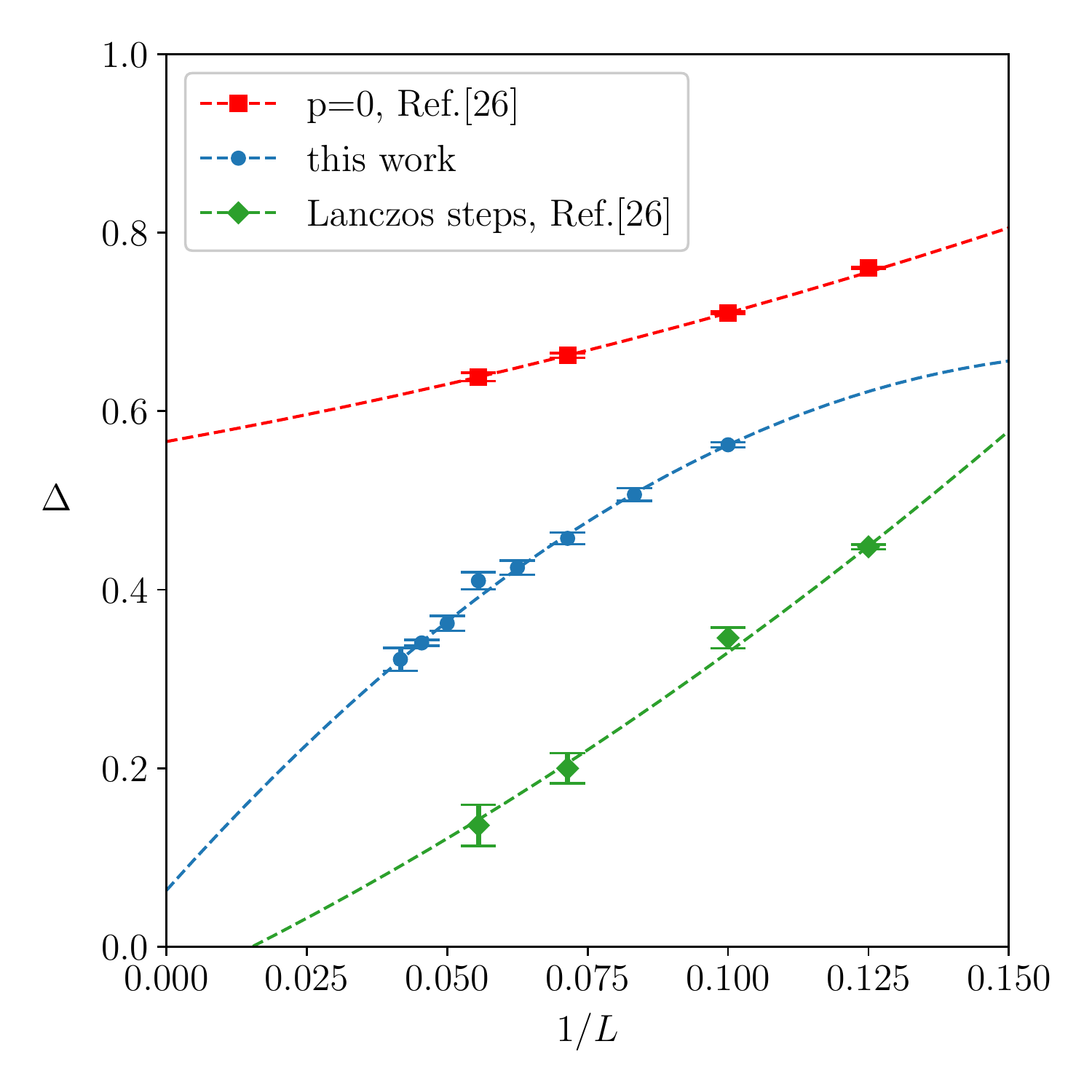}
\caption{
Finite-size scaling of the variational gap ($\Delta$) at $q=(\pi,0)$ within the spin-liquid phase ($J_2/J_1=0.55$). The results of Ref.~\cite{hu2013} are 
also reported for comparison. For the latter ones, the lowest-energy excitation with momentum $(\pi,0)$ is approximated by a single Gutzwiller-projected 
particle-hole excitation. The method employed in the present work makes use of a larger basis of excitations and, therefore, yields an improved finite-size 
scaling of the gap. The results obtained by a variance extrapolation with the Lanczos step procedure of Ref.~\cite{hu2013} is also reported for completeness.}
\end{figure*}

\begin{figure*}[!htb]
\includegraphics[width=0.5\columnwidth]{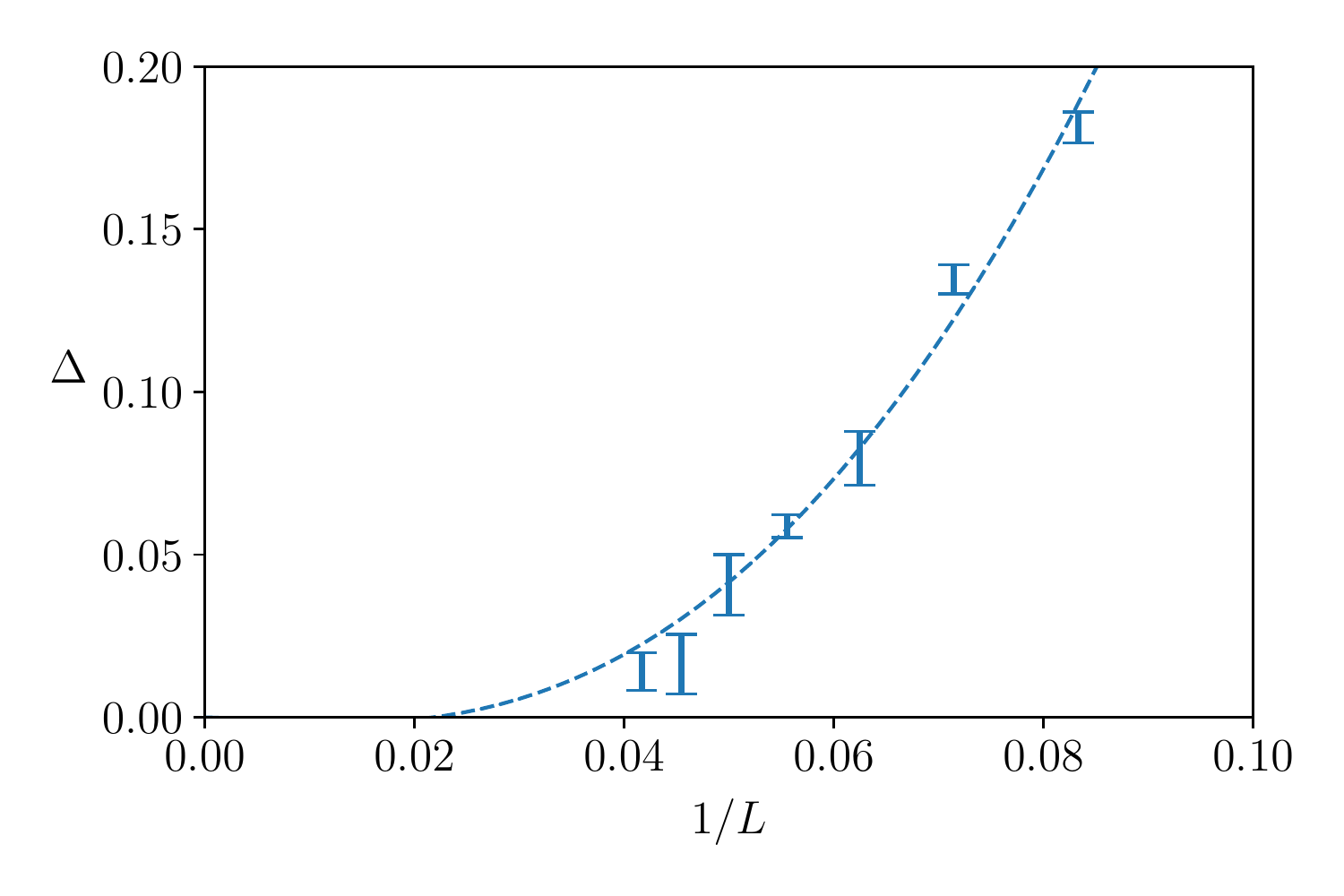}
\caption{
Finite-size scaling of the variational gap ($\Delta$) at $q=(\pi,\pi)$ within the spin-liquid phase ($J_2/J_1=0.55$). The value of the gap is rapidly 
decreasing to zero, suggesting a vanishing value in the thermodynamic limit. The dashed line is a guide to the eye.}
\end{figure*}

\end{document}